\documentclass[conference]{IEEEtran}

\pdfoutput=1

\usepackage{cite}
\usepackage[pdftex]{graphicx}
\usepackage{amsmath}
\usepackage{array}
\usepackage{url}

\usepackage{fancyhdr}   
\usepackage{amssymb}
\usepackage{amsthm}
\usepackage{mathtools}
\usepackage{color}
\usepackage{texdraw}
\usepackage{longtable}
\usepackage{colortbl}
\usepackage{lastpage}
\usepackage[T1]{fontenc}
\usepackage{algorithmic}
\usepackage{algorithm}
\usepackage{amsfonts}
\usepackage{listings}
\usepackage{color}
\usepackage{textcomp}
\usepackage{ulem}

\DeclareGraphicsExtensions{.pdf,.jpeg,.png}

\begin{document}

\normalem

\title{Integration of Satellites in 5G\\ through LEO Constellations}

\author{\IEEEauthorblockN{Oltjon Kodheli, Alessandro Guidotti, Alessandro Vanelli-Coralli}
\IEEEauthorblockA{Dept. of Electrical, Electronic, and Information Engineering (DEI), Univ. of Bologna, Bologna, Italy\\
Email: \{a.guidotti, alessandro.vanelli\}@unibo.it}\
oltjonkodheli@gmail.com

}

\maketitle

\begin{abstract}
The standardization of 5G systems is entering in its critical phase, with 3GPP that will publish the PHY standard by June 2017. In order to meet the demanding 5G requirements both in terms of large throughput and global connectivity, Satellite Communications provide a valuable resource to extend and complement terrestrial networks. In this context, we consider a heterogeneous architecture in which a LEO mega-constellation satellite system provides backhaul connectivity to terrestrial 5G Relay Nodes, which create an on-ground 5G network. Since large delays and Doppler shifts related to satellite channels pose severe challenges to terrestrial-based systems, in this paper we assess their impact on the future 5G PHY and MAC layer procedures. In addition, solutions are proposed for Random Access, waveform numerology, and HARQ procedures. 
\end{abstract}

\IEEEpeerreviewmaketitle

\section{Introduction}
During the last years, wireless communications have been experiencing an exploding demand for broadband high-speed, heterogeneous, ultra-reliable, secure, and low latency services, a phenomenon that has been motivating and leading the definition of new standards and technologies, known as 5G. The massive scientific and industrial interest in 5G communications is motivated by the key role that the future system will play in the worldwide economic and societal processes to support next generation vertical services, \emph{e.g.}, Internet of Things, automotive and transportation sectors, e-Health, factories of the future, etc., \cite{5GPPP_1,5GPPP_2}.\\
Due to the challenging requirements that 5G systems shall meet, \emph{e.g.}, large throughput increase, global and seamless connectivity, the integration of satellite and terrestrial networks can be a cornerstone to the realization of the foreseen heterogeneous global system. Thanks to their inherently large footprint, satellites can efficiently complement and extend dense terrestrial networks, both in densely populated areas and in rural zones, as well as provide reliable Mission Critical services. While in the past terrestrial and satellite networks have evolved almost independently from each other, leading to a difficult \emph{a posteriori} integration, the definition of the 5G paradigm provides the unique chance for a fully-fledged architecture. This trend is substantiated by 3GPP Service and system Aspects (SA) activities, which identified satellite systems as a possible solution both for stand-alone infrastructures and for complementing terrestrial networks, \cite{3GPPSA_1,3GPPSA_2}, as well as by EC H2020 projects as VITAL (VIrtualized hybrid satellite-TerrestriAl systems for resilient and fLexible future networks), in which the combination of terrestrial and satellite networks is addressed by bringing Network Functions Virtualization (NFV) into the satellite domain and by enabling Software-Defined-Networking (SDN)-based, federated resources management in hybrid SatCom-terrestrial networks, \cite{VITAL}.

In this context, the integration of terrestrial systems with Geostationary Earth Orbit (GEO) satellites would be beneficial for global large-capacity coverage, but the large delays and Doppler shifts in geostationary orbits pose significant challenges. In \cite{Intro1,Intro2,Intro3}, resource allocation algorithms for multicast transmissions and TCP protocol performance were analyzed in a Long Term Evolution (LTE)-based GEO system, providing valuable solutions. However, to avoid the above issues, significant attention is being gained by Low Earth Orbit (LEO) \emph{mega-constellations}, \emph{i.e.}, systems in which hundreds of satellites are deployed to provide global coverage, as also demonstrated by recent commercial endeavors.  In \cite{Intro_UniBo}, we considered a mega-constellation of LEO satellites deployed in Ku-band to provide LTE broadband services and analyzed the impact of large delays and Doppler shifts on the PHY and MAC layer procedures.

Since 3GPP Radio Access Network (RAN) studies and activities are now providing significant results and critical decisions have been made during the last meetings on the PHY and MAC layers for the New Radio (NR) air interface, which will be finalized by June 2017, \cite{IEEEexample:UEmob,3GPP_SI}, it is of outmost importance to assess the impact that these new requirements will have on future 5G Satellite Communications (SatCom). To this aim, in this paper, we move from the analysis performed in \cite{Intro_UniBo} and assess the impact of large delays and Doppler shifts in a LEO mega-constellation operating in Ku-band. In particular, we will focus on waveform design, Random Access, and HARQ procedures.

\section{System Architecture}
\begin{figure}[!t]
\centering
\includegraphics[width=0.45 \textwidth]{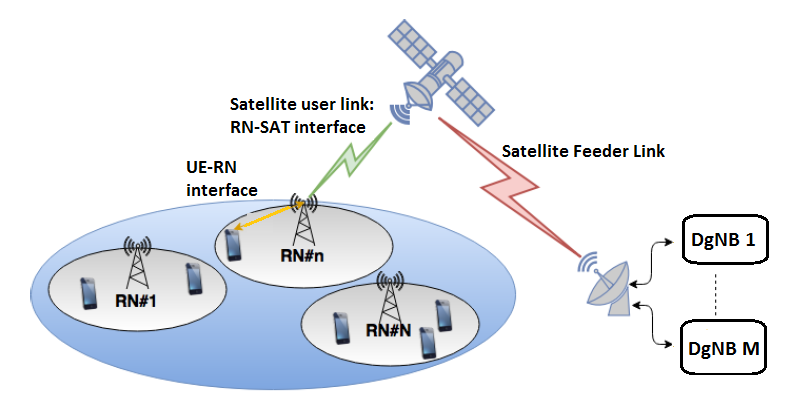}
\caption{g-Sat-RN architecture \cite{Intro_UniBo}}
\label{fig_sim1}
\end{figure}

We consider a system architecture similar to that addressed in \cite{Intro_UniBo}, with a mega-constellation of Ku-band LEO satellites, deployed at $h=1200$ km from Earth, beam size set to $320$ km, and a minimum elevation angle of $45$ degrees. In the proposed heterogeneous architecture, we introduce on-ground Relay Nodes (RN) that extend the terrestrial network coverage with reduced costs. RNs were introduced and standardized in LTE, \cite{IEEEexample:rnstandard}, and within the 3GPP standardization framework it is agreed that they will be also implemented in 5G, \cite{IEEEexample:rn5G}.\\
The following assumptions are made for the proposed system architecture, which is shown in Fig.~\ref{fig_sim1}: i) the terrestrial terminals are 5G User Equipments (UE) connected to the 5G RNs that provide the terrestrial access link; ii) we assume the deployment of Type-1 RNs, which have up to layer-3 capabilities, \emph{i.e.}, they can receive, demodulate, and decode data, apply another FEC, and then re-transmit a new signal. They also have their own cell ID, synchronization, broadcast, and control channels; iii) the satellites are assumed to be transparent and to provide backhaul connectivity to the on-ground RNs; and iv) the satellite gateway is connected to the satellites through ideal feeder links, providing access to the Donor gNB (DgNB) that connect the RN with the 5G Core Network. It is worthwhile to highlight that the RNs act as gNB (5G NodeB, as per 3GPP nomenclature) from the User Equipment (UE) perspective and as UE when communicating with a donor gNB (DgNB).\\
The focus of this paper is on the backhaul link between the RN and the DgNB, and in particular on the impact of large delays and Doppler shifts on the PHY/MAC procedures. We assume the link to be a modified version of UE-RN interface, which will have the same protocols, but with different RF characteristic. For instance, it is already agreed that the 5G air interface will use Ciclic Prefix-Orthogonal Frequency Division Multiplexing (CP-OFDM) waveforms, coupled with scalable numerology, and UE-RN interface can use different numerologies with respect to RN-DgNB interface. \\
We assume a Frequency Division Duplexing (FDD) frame structure, since Time Division Duplexing (TDD) is not suitable to scenarios affected by large delays. Finally, outband RNs are assumed, \emph{i.e.}, backhaul and the access links use different frequencies. This isolation is important, because the generic UE will covered by both the related RN and the satellite, which might cause interference without a proper frequency separation. It is worth noting that one DgNB can be connected with more than one RN and $M\leq N$, where $M$ and $N$ represent the number of DgNBs and RNs respectively.

\section{Satellite Channel}

In this section, we focus on the main satellite channel impairments, \emph{i.e.} large Round Trip Time (RTT) and Doppler shifts, in order to assess their impact on the 5G MAC and PHY procedures.
\subsection{Delay}

In order to compute the RTT, we have to take into account the distance from RN to the satellite, and the distance from the satellite to the DgNB. There should be also the path between UEs and RNs, but it is negligible compared to the other terms. The minimum distance between the satellite and the RN is reached with an elevation angle of $90$ degrees and is equal to $h=1200$ km. By approximating also the distance between the satellite and the DgNB $d_{S-D}$ with the altitude $h$, we have:
\begin{equation}
RTT = 2*\frac{h+d_{S-D}}{c}\approx 4*\frac{h}{c}\approx 16 \ \mathrm{ms}
\end{equation}
where $c$ is the speed of light. This value of is quite large if compared with the maximum RTT foreseen in a terrestrial network. This is a critical challenge for integrating satellites with 5G systems and will be addressed in the following sections.

\subsection{Doppler}

In order to compute the Doppler shift, two separate scenarios have to be treated. First, we consider the Doppler shift that the UE experiences due to its relative motion to RN. If $v$ denotes the speed of the UE relative to the RN, $f_c$ the carrier frequency, and $\theta$ the angle between velocity vectors, the maximum Doppler frequency is given by:
\begin{equation}
f_{d}=\frac{v*f_c}{c}*cos(\theta)
\end{equation}
The targeted mobility in 5G is $500$ km/h for carrier frequencies below 6 GHz, \cite{IEEEexample:UEmob}. This is the maximum UE speed at which a minimum predefined quality of service (QoS) can be achieved. Therefore, at $4$ GHz carrier frequency, supporting a $500$ km/h mobility for the UE would lead to a Doppler spread up to $1.9$ kHz. Furthermore, transmissions in above $6$ GHz bands will be more sensitive to frequency offset.
It means that NR will be able to correctly estimate and decode the signal, even though a Doppler shift of $1.9$ kHz is experienced.

Second, we compute the Doppler shift experienced on the satellite channel. This Doppler shift will be caused only by the movement of satellite, because both the RN and the DgNB are fixed on Earth. A closed-form expression for the Doppler in LEO satellites is provided in \cite{Intro_UniBo} and is as follows:
\begin{equation}
\label{eq:equation1}
f_d(t)=\frac{f_c*w_{SAT}*R_E*cos(\theta(t))}{c}
\end{equation}
where $w_{SAT}$ is the angular velocity of the satellite $w_{SAT}=\sqrt[]{G*M_E/(R_E+h)^3}$, $R_E$ is the Earth radius, $\theta(t)$ is the elevation angle at a fixed time $t$, $M_E$ is the mass of Earth, and $G$ the gravitational constant. In the considered scenario, we can derive the maximum Doppler shift to be $11\ \mathrm{GHz} < f_d < 14\ \mathrm{GHz}$ in Ku-band. Clearly, this value is significantly larger than that experienced in a terrestrial link.

\section{Challanges of 5G air interface (NR) in satellite context}

In this section, we assess the impact of the large RTT and Doppler shifts the NR MAC and PHY layers. There is an ongoing work towards 5G standardization, including MAC and PHY layer aspects. Some significant high-level agreements have been reached in the latest 3GPP meetings. On the basis, we highlight the challenges that the satellite link imposes to NR, in order to be ``satellite friendly.''

\subsection{Waveform}

Three scenarios have been defined by ITU-R for IMT-2020 (5g) and beyond \cite{IEEEexample:imtu}: i) eMBB (enhanced Mobile BroadBand); ii) mMTC (massive Machine Type Communications); and iii) uRLLC (Ultra-Reliable and Low Latnecy Communications). It is agreed that CP-OFDM waveforms are supported for eMBB and uRLLC use cases, while Discrete Fourier Transform-spread-OFDM (DFT-s-OFDM) waveforms are also supported at least for eMBB uplink for up to $40$ GHz \cite{IEEEexample:3gpp,IEEEexample:3gpp2}. With respect to the LTE waveforms, the NR waveforms are more flexible. This flexibility is given by different sub-carrier spacing (SCS) and filtering/windowing techniques. Depending on the use case, NR will use different numerologies across different UEs. Scalable numerology should allow at least from $15$ kHz to $480$ kHz subcarrier spacing, \cite{IEEEexample:techrep22}. Basically, the values of SCS can be defined as: $SCS = 15 kHz * 2^n$, where $n$ is a non negative integer.\\
It is straightforward to say that larger SCS lead to increased robustness to Doppler shifts. Depending on the UE speed and the carrier frequency used, SCS should be large enough to tolerate the Doppler. For example, in LTE, with a $2$ GHz carrier frequency, the maximum Doppler that can be supported is $950$ Hz because the SCS is fixed to $15$ kHZ. Basically, it corresponds to $6.3 \%$ of the SCS. NR can cope with larger Doppler shift values because it can adopt SCS larger than $15$ kHz. As outlined in the previous section, NR should support up to $1.9$ kHz of Doppler shift for a $4$ GHz of carrier frequency. This is totally possible, by using SCS of $30$ kHz ($6.3\%*30 \ \mathrm{kHz} \approx 1.9 \ \mathrm{kHz}$). When going above $6$ GHz, the sensitivity to frequency offset is larger and higher SCS can be implemented. 

The Doppler shift in the considered satellite system is still very large even in when relying on the highest SCS made available by NR ($480$ kHz), which corresponds to a value of $30.4$ kHz of tolerated frequency offset. 
It is worth noticing that, in our scenario, the only interface affected by high Doppler through satellite is the RN-DgNB interface. The Doppler experienced due to UE movement (high speed trains), is estimated and compensated by RNs. Therefore, no modification is needed in the interface UE-RN. Solutions should be found for RN-DgNB interface, in order to cope with extremely high Doppler in the satellite link.

\subsection{Random Access}

The Random Access (RA) procedure is used by the UEs to synchronize with the gNB and to initiate a data transfer. During 3GPP RAN meetings, the following high level agreements were made about RA procedure in NR, \cite{IEEEexample:3gppra2}: i) both contention-based and contention-free RA procedure should be supported in NR; and ii) contention-based and contention-free RA procedure follow the steps of LTE. Basically, contention-based RA is performed for initial access of the UE to the gNB, or for re-establishment of synchronization in case it is lost. Whereas, contention-free RA is performed in handover situation, when the UE was previously connected to a gNB.\\
There are four steps for contention-based RA procedure in NR. Steps 1 and 2 mainly aim at synchronizing uplink transmission from UE to gNB. The UE randomly chooses a preamble from a predefined set and sends it to the gNB. The preamble, consists of a cyclic prefix of length $T_{cp}$, a sequence part of length $T_{SEQ}$, and a guard time $T_G$ in order to avoid collision of RA preambles. The gNB receives the preamble and sends a RA Response (RAR) to the UE with information about timing advance (TA), a temporary network identifier (T-RNTI), and the resources to be used in step 3. For the moment being, there is no agreement on the size of RAR window in NR.\\
In steps 3 and 4, a final network identifier CRNTI is assigned to the UE. In LTE, the contention timer in which the UE receives the final network identifier can be as large as $64$ ms. In NR there is no agreement so far related to this value. \\
In our scenario we have taken into account the presence of RN, so we have to analyze the RA procedure in two stages, as follows.
\subsubsection{UE random access procedure}
Each UE should perform the RA procedure with the corresponding RN. It is worth noticing that contention-free RA involves only steps 1 and 2. In these first two steps, the RN does not need to contact the DgNB. It can terminate all protocols up to Layer 3. The delay in the satellite link is not involved, therefore contention-free RA procedure can be implemented without modifications. \\
On the other hand, in contention-based RA, in order to provide a final network identifier to the UEs (step 3 and 4), the RN should contact the Core Network through DgNB. In this case, the RTT time on the satellite channel should be considered. The contention timer in NR should be larger than $16$ ms, in order to support RA procedure through satellite link. In LTE, there is a contention timer up to $64$ ms. Even though, the tendency is to have lower value in NR, no drastic reduction is expected. Hence, UE-RN RA procedure is expected to be realizable without any modification. 
\subsubsection{RN attach procedure}
When a new RN is installed in the network, it should automatically begin its start-up attach procedure, which is performed in two phases,\cite{IEEEexample:techrep}. In the first phase, in order to have information for the initial configuration, RN attaches to the network as a UE performing the same steps of RA procedure. In the second phase, the RN receives from DgNB further specification (\emph{e.g.}, IP address of the S-GW/P-GW), so as to be able to operate as a relay.
In contrast with the UE RA procedure, here the RTT of satellite channel is present in all steps. Therefore, not only the contention timer should be more than $16$ ms, but also the RAR window size. In LTE, the RAR window size is between $3$ and $15$ ms and lower values are expected for NR \cite{IEEEexample:techrep4}. Possible solutions should be found about the RAR window size.

\begin{figure}[!t]
\centering
\includegraphics[width=0.38 \textwidth]{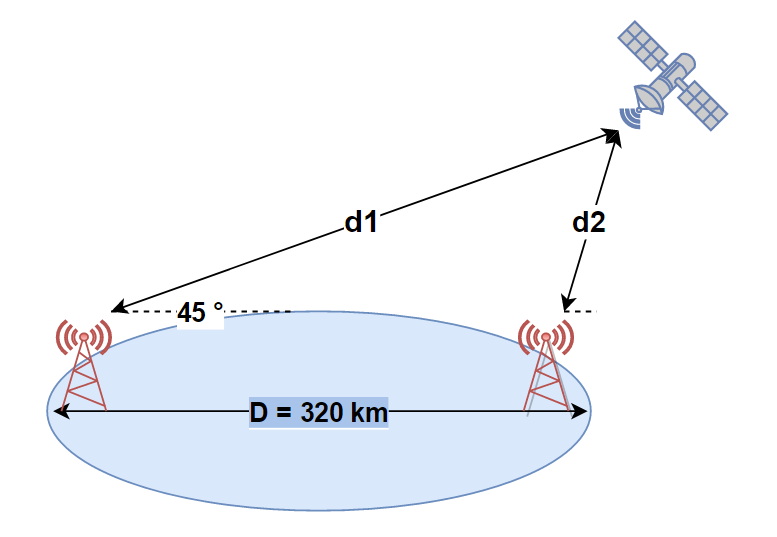}
\caption{Two RN with the largest possible distance}
\label{fig_sim6}
\end{figure}

Another thing to be considered, is the preamble format of the RA procedure for RNs. The length of the preamble in the RA procedure defines the coverage that can be supported by the standard. In LTE there are $4$ preamble formats, and the largest one is designed to support a $100$ Km radius cell, corresponding to $0.67$ ms Timing Advance (TA), \cite{IEEEexample:techrep3}. In NR, at least for deployments below $6$ GHz, longer RA preambles should be supported, \cite{IEEEexample:3gppreamble}. It means that NR should support at least the same coverage as LTE. Therefore, we have to calculate the TA needed in our scenario for RN attach procedure through satellite link. To do so, we have to find the difference in RTT between $2$ RNs, one with the shortest and the other one with the longest path to the satellite. The worst scenario, where the differences of RTT between two RNs is at its maximum occurs at minimum elevation angle, which in our case is $45$ degrees, as shown in Fig.~\ref{fig_sim6}.
By geometrical considerations, we can find that the maximum allowable distance between the RN and the satellite is $d_1 = 1580$ km, by solving the following equation:
\begin{equation}
(R_E+h)^2 - R_E^2 - |d_1|^2 = 2R_E|d_1|sin(\theta)
\end{equation}
where $\theta$ is the minimum elevation angle. As the beam has a diameter of $320$ km, this is also the max distance between $2$ RNs. It can be easily computed that $d_2 = 1372$ km, which results to $|d_1 - d_2|= 208$ km. This value is almost twice larger than the cell size that can be supported in NR. Valuable solutions should be found.

Finally, it is important to mention that at start-up deployment of network, TA and RAR window size does not pose any challenge in RN attach procedure, because all the RNs can be introduced in the network one-by-one. However, under certain circumstances (\emph{e.g.}, satellite handover, heavy rain, etc.) RNs can loose their synchronization to DgNB, therefore a repetition of RN attach procedure is needed and all the RNs will compete for the channel at the same time.

\subsection{Satellite Handover}

LEO satellites move at high speeds above the ground. Even though no mobility is foreseen for RNs, due to the movement of the satellite, RNs will be seen at a certain amount of time. Therefore, all the system should pass through handover process. It means that each RN should perform RA procedure periodically, in order to obtain the physical link with the DgNB through another satellite. Performing a RA procedure periodically for RNs is very time-expensive. Consequently, there can be a significant reduction of throughput. Other solutions should be found. A possible solution can be found in  \cite{Intro_UniBo}.

\subsection{Hybrid Automatic Repetition Quest (HARQ)}

\begin{figure}[!t]
\centering
\includegraphics[width=0.45 \textwidth]{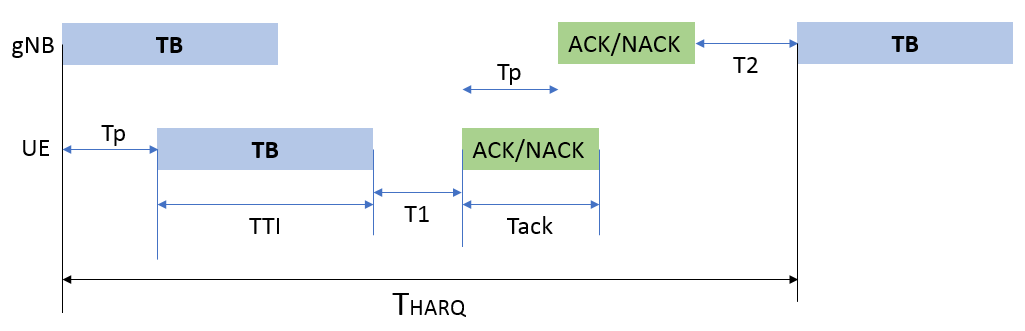}
\caption{Timing diagram of one HARQ process}
\label{fig_sim2}
\end{figure}

One of the main requirements for 5G is to improve the link reliability. To this aim, among other solutions, HARQ protocols will be implemented as in LTE, \cite{IEEEexample:techrep55}. If the transmission block (TB) is decoded correctly at the receiver, it responds with an ACK, otherwise it will send a NACK and the packet will be re-transmitted by adding more redundancy bits. In order to improve system efficiency, multiple parallel HARQ processes are used. That means that, in order to be able to achieve peak data rate, the following formula should hold:
\begin{equation}
N_{HARQ,min}\geq \frac{T_{HARQ}}{TTI} 
\end{equation}
where $N_{HARQ,min}$ is the minimum number of HARQ processes, $TTI$ is the transmission duration of one TB, and $T_{HARQ}$ is the time duration between the initial transmission of one TB and the corresponding ACK/NACK.
This is illustrated in Fig.~\ref{fig_sim2}.
$T_1$ and $T_2$ are the processing time at the UE and gNB respectively, $T_{ack}$ is the TTI duration of ACK/NACK and $T_p$ is the propagation time from gNB to UE or vice versa.\\
Based on these considerations, in FDD LTE up to $8$ parallel HARQ processes are used both in the downlink and the uplink. The numerology is fixed, hence there is just one HARQ process configuration. It is agreed that NR \cite{IEEEexample:3gpp}\cite{IEEEexample:3gpp2} should be able to support  more than one HARQ processes for a given UE and one HARQ process for some UEs. It means that, depending on the use case, different HARQ configurations can be supported. This gives us a desired degree of freedom, when considering HARQ procedure over a sat-channel.
 \\
In our scenario, the critical issue is the high RTT ($16$ ms) in the satellite link, which will have a significant impact in HARQ process, especially to the overall time $T_{HARQ}$. In a terrestrial link, the impact of RTT is negligible to HARQ, because the processing time at gNB/UE is much larger than the propagation delay. Whereas, in the considered LEO system, RTT now has a major role in the HARQ protocol. If we consider a total processing time of 8 ms (like in LTE) and a $1$ ms TTI we can calculate:
\begin{equation}
N_{HARQ,min}^{SAT}\geq \frac{T_{HARQ}}{TTI}=\frac{16 + 8 }{1} =24 
\end{equation}
As we can see, NR should be able to support $24$ parallel HARQ processes in order to be compatible and to offer the desired peak data rate over a LEO satellite link. \\
The possible large number of HARQ processes would have an impact on the following: i) soft-buffer size of the UE ($N_{buffer} \propto N_{HARQ,max}*TTI$); ii) bit-width of DCI fields (3 in LTE because there are 8 processes). On one hand, increasing UE buffer size can be very costly and on the other hand larger bit-widths of DCI field would lead to large DL control overhead. These issues should be treated carefully and solutions must be found.

\section{Proposed Solutions}

\begin{figure}[!t]
\centering
\includegraphics[width=0.35\textwidth]{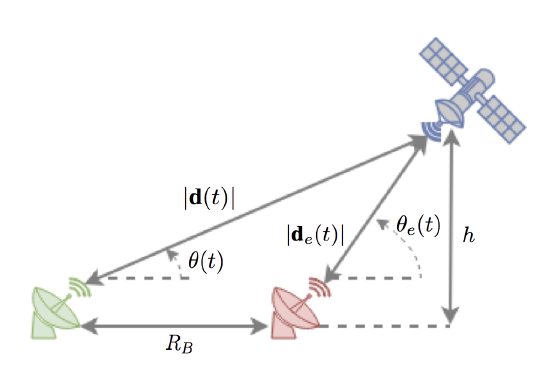}
\caption{RN position displacement \cite{Intro_UniBo}}
\label{fig_sim3}
\end{figure}

\subsection{Waveform}

\begin{figure}[!t]
\centering
\includegraphics[width=0.4 \textwidth]{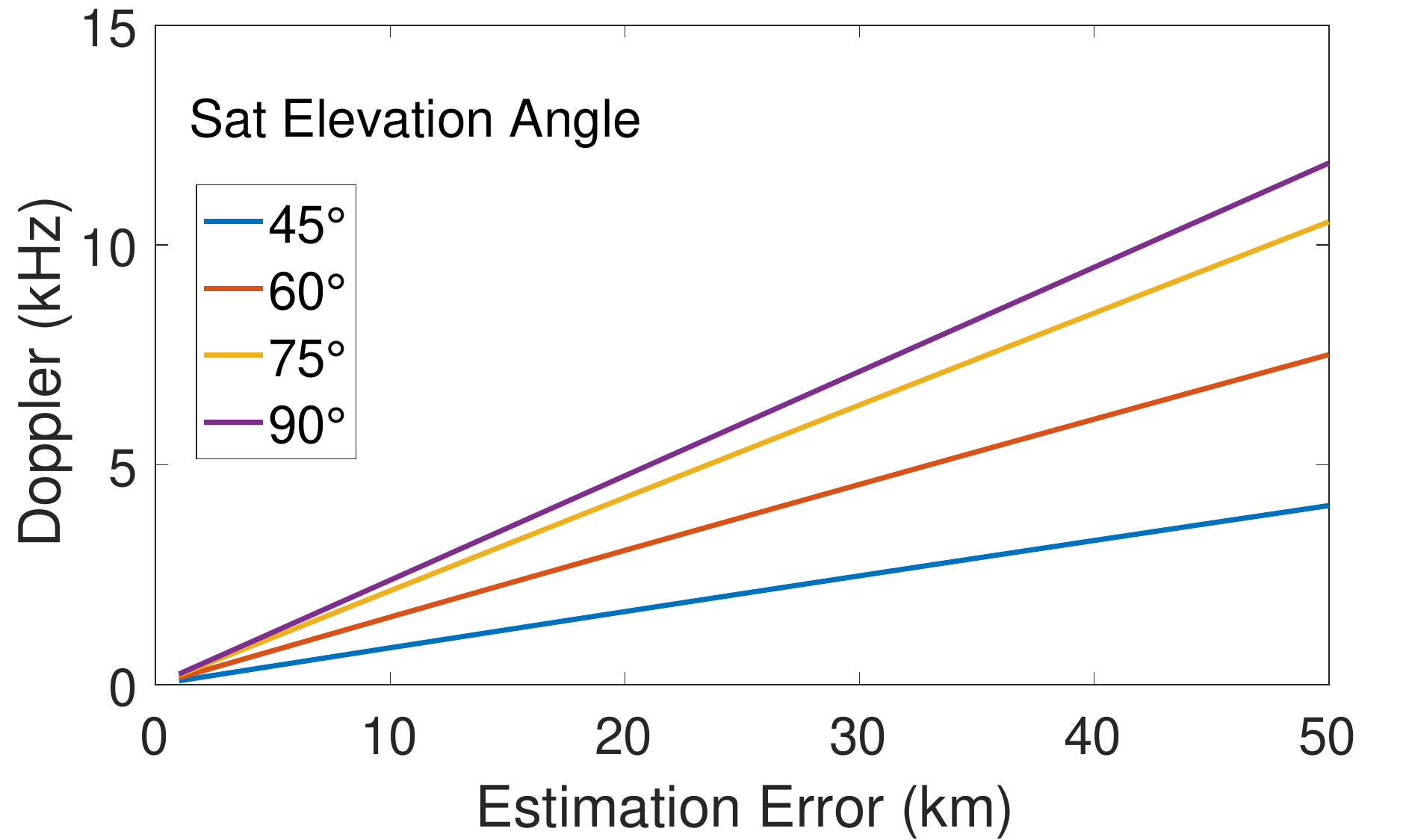}
\caption{Doppler as a function of position estimation error}
\label{fig_sim}
\end{figure}

In order to deal with large Doppler shifts in the satellite channel, we have to find valuable solutions without changing 5G waveforms to assure forward compatibility. We propose to equip RNs with GNSS receivers, being able to estimate the position of the satellite. By doing so, we can compensate the Doppler shift significantly. For sure, whenever we deal with an estimator, we will have some error characterized by a variance of error. Due to the estimation error, a residual Doppler shift will be generated. 
Let us characterize the residual Doppler as a function of estimation error as follows: \\
Given the position of the satellite $X_{SAT}(t)$ at a certain time, we can calculate the distance $|d(t)|$ and the elevation angle, $\theta(t)$ by simple geometry, as shown in Fig.~\ref{fig_sim3}. By substituting this value of elevation angle, we obtain the real Doppler shift. However, due to an error in estimating the position ($R_B$), we will have a new duplet ($|d_e(t)|$,$\theta_e(t)$). These values can be related to the real duplet ($|d(t)|$,$\theta(t)$) as follows:
\begin{equation}
|d_e(t)|*cos(\theta_e(t))=|d(t)|*cos(\theta(t)) + R_B 
\end{equation}
\begin{equation}
|d(t)|^2=|d_e(t)|^2+R_B^2 - 2*R_B*|d_e(t)|*cos(\theta_e(t)) 
\end{equation}
By solving (7) and (8) we obtain:
\begin{equation}
cos(\theta_e(t)) = \frac{|d(t)|*cos(\theta(t))+R_B}{\sqrt{|d(t)|^2+R_B^2 + 2*R_B*|d(t)|*cos(\theta(t))}}
\end{equation}
By plugging this result to equation~\ref{eq:equation1}, we finally obtain the residual Doppler $\tilde{f_d(t)}$. The result is illustrated in Fig.~\ref{fig_sim} for values of $0\leq R_B\leq 50$ and $45\leq \theta \leq 90$. The results are summarized in Table 1.
\\
It is worth mentioning that the highest Doppler is experienced at 90 degrees of elevation angle under a fixed estimation error.
we can  conclude that,  increasing the SCS, larger estimation errors can be tolerated. Therefore, it is preferable a waveform with high SCS in RN-DgNB interface. 

\subsection{RA Procedure}

For the RA procedure, the RAR window size and TA in case of RN-DgNB RA shall be considered.

\subsubsection{RAR window size} Two solutions can be identified.

Option 1: NR should fix a value of the RAR window, taking into account the worst scenario, having the largest delay. Hence RAR window size should be larger than $16$ ms for all use cases. This will allow to deal with large delays in a satellite channel, but will impose useless high delays in the network, where a terrestrial channel is used.

Option 2: RNs, before starting the RA procedure, should be informed from their respective downlink control channel (R-PDCCH), about the presence of a satellite link. Depending on this, the RNs should be able to change the RAR window size. Clearly, adding more bits in the PDCCH will increase the overhead, but in this case it is needed just one bit more to indicate the presence of a sat-channel or not.
 
\subsubsection{Taming advance (TA)} Two solutions are proposed.

Option 1: the length of preamble format is also the indication of the maximum supported cell size. NR should use larger preamble formats in order to cope with larger TA due to satellite channel. A detailed analysis is needed about the preamble format.

Option 2: by equipping RNs with GNSS receivers, they can estimate the satellite position and their distance. Therefore, it is possible to estimate the TA before sending the preamble to DgNB. By doing so, there is no need to change the preamble format. Due to the estimation error, RNs with receive the exact TA in the RAR message, which is going to be much lower. An estimation error up to $100$ km (corresponding to the largest cell size) can be supported in order to keep the same preamble format. 

\subsection{HARQ Retransmissions}

In order to deal with the large possible number of HARQ re-transmission, 4 solutions are identified.
\subsubsection{Increased number of HARQ processes and buffer size}
As we highlighted in the previous section, for assuring continuous transmission and achieving peak data rate, we need at least $24$ parallel HARQ processes. The increase on the buffer size and whether the UE will have this capability of memory, need further analysis.
\subsubsection{Increased number of HARQ processes with reduced buffer size}
It is possible to increase the number of HARQ processes, by maintaining the buffer size under control. In LTE, we reserve 1 bit for ACK/NACK. By enhancing the feedback information using 2 bits  \cite{multibit}, we can inform the transmitter on how close the received TB is to the originally TB. Therefore, the number of re-transmission will be reduced, because the transmitter can add the redundant bits according to the feedback information. Reducing the number of re-transmission has a direct impact on the buffer memory size.
\subsubsection{Reduced number of HARQ processes with reduced buffer size}
In case the first two solutions are not possible due to UE capability, another solution would be to keep the number of HARQ processes under a certain limit. It is worth noticing that the throughput of transmission will be reduced.
\subsubsection{No HARQ protocol}
In case HARQ in not supported due to large delay in the satellite channel, a solution would be to replicate the TB a certain amount of times before sending to the receiver. This will reduce the impact of the delay, but the throughput will be reduced significantly.

\begin{table}[!t]
\renewcommand{\arraystretch}{1.3}
\caption{Maximum poition error tolerated in different SCS}
\label{table_example}
\centering
\begin{tabular}{c||c||c}
\hline
\bfseries SCS (kHz) & \bfseries Max. Doppler (kHz)& \bfseries Max. position error (km)\\
\hline\hline
15 & 0.95 & 3.95\\
\hline\hline
30 & 1.9 & 7.9\\
\hline\hline
60 & 3.8 & 15.8\\
\hline\hline
120 & 7.6 & 31.6\\
\hline
\end{tabular}
\end{table}

\section{Conclusions}
In this paper, we proposed a 5G-based LEO mega-constellation architecture to foster the integration of satellite and terrestrial networks in 5G. By referring to the latest 3GPP specifications, we addressed the impact of typical satellite channel impairments as large delays and Doppler shifts on PHY and MAC layer procedures, as well as on the waveform numerology. The impact of the Doppler shift on the waveform and the effect of large delays in RN-DgNB RA procedure, can be estimated and compensated by accurate GNSS receivers.  Increasing SCS in the backhaul link, allows larger tolerated estimation error. We also propose that RNs should be informed by control channel (R-PDCCH) about the presence of the satellite link, in order to increase RAR window size.
To achieve the peak data rate, at least 24 HARQ parallel processes are needed, posing a big challenge on UE buffer size and DCI field. Some solutions have been proposed for keeping the number of HARQ processes and buffer size under control.

\end{document}